\documentstyle[prd,aps,preprint,psfig]{revtex}

\begin{document}

\draft

\preprint{OKHEP-96-08}
\title{Analytic perturbation theory in QCD and Schwinger's
connection between the $\beta$-function and the spectral density}
\author{K. A. Milton$^a$\thanks{E-mail: milton@phyast.nhn.ou.edu}
 and I. L. Solovtsov$^{ab}$\thanks{E-mail:
solovtso@thsun1.jinr.dubna.su}}
\address{$^a$Department of Physics and Astronomy,
         University of Oklahoma, Norman, OK 73019 USA\\
         $^b$Bogoliubov Laboratory of Theoretical Physics,
         Joint Institute for Nuclear Research,\\
         Dubna, Moscow Region, 141980, Russia}
\date{\today}

\maketitle
\begin{abstract}

We argue that a technique called analytic perturbation theory
leads to a well-defined method for analytically continuing
 the running coupling constant from the spacelike to the
timelike region, which allows us to give a self-consistent
definition of the running coupling constant
for timelike momentum. The corresponding $\beta$-function
is proportional to the  spectral density, which confirms a
hypothesis due to Schwinger.
\end{abstract}


\section{Introduction}

An outstanding problem in QCD is the extrapolation of limited
perturbation-theory information so as to make contact with
experiment.
An important example is given by the running coupling constant, in
which
low-order calculations are summed by the renormalization group.  It
has
been known since the early 1950's that this is really not
self-consistent,
because of the appearance of an unphysical spacelike singularity, the
``ghost-pole.'' However, as it has been argued in
Refs.~\cite{ss1,ss2},
a possible way to resolve the ghost-pole  problem  for the QCD
running
coupling constant can be found by imposing  K\"all\'en--Lehmann
analyticity.
This method, which was elaborated early in the development of QED
\cite{red,trio} leads to the definition of the analytic
running coupling constant in the complex $q^2$-plane with a cut along
the negative part of the real axis. (In this paper we use a metric
with
signature $(-1,1,1,1)$, so that $q^2>0$ corresponds to a spacelike
momentum
transfer.) According to Refs.~\cite{ss1,ss2} the connection between
the analytic coupling $\bar{a}(q^2)$ and the spectral density
$\rho (\sigma)$ is given by the following spectral representation
(the overbar signifies the analytically improved quantity)
\begin{equation}
\label{spectral}
\bar{a}(q^2)\equiv\frac{\bar{\alpha}_s(q^2)}{4\pi}=\frac{1}{\pi}
\int_0^\infty d\sigma \frac{\rho (\sigma)}{\sigma+q^2-{\rm
i}\epsilon} .
\end{equation}
(The questions about the validity of the spectral representation,
which is rather obvious in QED, are resolved in \cite{ginzburg} for
the
general case.)
For instance, in the one-loop approximation to the  spectral
function,
\begin{equation}
\rho^{(1)}(\sigma)={\rm Im}\,\bar a^{(1)}(-\sigma-{\rm i}\epsilon)
={\rm Im}\, {a\over1+a\beta_0\ln(-\sigma/\mu^2-{\rm i}\epsilon)}
={a^2\beta_0\pi\over[1+a\beta_0\ln(\sigma/\mu^2)]^2
+a^2\beta_0^2\pi^2},
\label{specfn}
\end{equation}
the corresponding analytic running coupling constant has the form
\begin{equation}
\label{a1}
\bar{a}^{(1)}(q^2)= \frac{1}{\beta_0}
\left[\frac{1}{\ln q^2/\Lambda^2}\,+\,\frac{\Lambda^2}
{\Lambda^2-q^2}\right],
\end{equation}
where $\beta_0=11-2/3\, N_f$ is the first coefficient of the
$\beta$-function
with $N_f$ active flavors, and $\Lambda$ is the QCD scale.
The analytically-improved coupling constant (\ref{a1}) has
no ghost pole at $q^2=\Lambda^2$, and its correct analytic properties
are
provided by the nonperturbative contribution, the second term in
(\ref{a1}),
which has appeared automatically through use of the spectral
representation (\ref{spectral}). The analytic running coupling
constant
 obtained in such a way turns out to be remarkable stable in the
 infrared region with respect to higher loop corrections and
 has the universal infrared limit at
$q^2=0$: $ \bar{a}(0)=1/\beta_0$, which does not depend on the
value of $\Lambda$, being a universal constant.

\section{Spacelike and timelike running couplings}

The method described above defines the running coupling constant in
the
Euclidean (spacelike) range of momentum, $q^2>0$, where $\overline
a(q^2)$
is real.  In this paper we wish to parametrize processes with
timelike momentum transfer, for example, the process of $e^+e^-$
annihilation into hadrons.  To do so, we must make use of
some nontrivial analytic continuation procedure from the spacelike to
the timelike region. To this end one usually applies the dispersion
relation for the Adler $D$-function, defined in terms of the
correlation function for the quark vector current, $\Pi(q^2)$, as
follows
\begin{equation}
\label{dq2p}
D(q^2)=-q^2\frac{d\Pi(-q^2)}{ dq^2}.
\end{equation}
This ``vacuum polarization'' satisfies an unsubtracted dispersion
relation,
\begin{equation}
\Pi(-q^2)={\rm const.}+\int_0^\infty {ds\over s+q^2}R(s),
\label{pi}
\end{equation}
with the $e^+e^-$ annihilation ratio is  given by
\begin{equation}
\label{rsp}
R(s)=\frac{1}{2\pi{\rm i}}
\left[\Pi (s+{\rm i}\epsilon)-\Pi (s-{\rm i}\epsilon)\right].
\end{equation}
Consequently, the dispersion relation for the Adler function is
\begin{equation}
\label{dq2}
D(q^2)=q^2\int_0^{\infty}\frac{ds}{{(s+q^2)}^2}R(s),
\end{equation}

The $D$-function is an analytic function in the complex $q^2$ plane
with a cut along the negative real axis. Taking into account
these analytic properties we can write down the inverse relation for
$R(s)$,
\begin{equation}
\label{rs}
R(s)=-\frac{1}{2\pi{\rm i}}
\int _{s-{\rm i}\epsilon}^{s+{\rm i}\epsilon} \frac{dz}{z}D(-z),
\end{equation}
where the contour goes from the point
$z= s-{\rm i}\epsilon$ to the point $z= s+{\rm i}\epsilon$
and lies in the region of analyticity of the function $D(z)$.

Let us define effective coupling constants $\bar{a}^{\rm eff} (q^2)$
in the spacelike region and $\bar{a}^{\rm eff}_s(s)$ in the timelike
region based on the following expressions for $D(q^2)$ and $R(s)$
\begin{equation}
\label{dqeff}
D(q^2)\propto
\left[1+d_1\bar{a}^{\rm eff}(q^2)\right],
\end{equation}
\begin{equation}
\label{rqeff}
R(s)\propto
\left[1+r_1{\bar{a}}_s^{\rm eff}(s)\right],
\end{equation}
where $d_1$ and $r_1$ are the first coefficients of perturbative
expansions.
(The superscript $\rm{eff}$ refers to the summation of all the
remaining
terms in the perturbative expansion of these quantities.)
In fact, $d_1=r_1$.
The subscript $s$ in (\ref{rqeff}) means ``$s$-channel"
(the timelike region).
{}From (\ref{dq2}) and (\ref{rs}),
one finds the connections between these effective coupling constants
in
the spacelike and timelike regions:
\begin{equation}
\label{aeff}
\bar{a}^{\rm eff}(q^2)=q^2\int_0^{\infty}
\frac{ds}{{(s+q^2)}^2}\bar{a}_s^{\rm eff}(s)
\end{equation}
and
\begin{equation}
\label{aseff}
\bar{a}_s^{\rm eff}(s)=-\frac{1}{2\pi {\rm i}}
\int _{s-{\rm i}\epsilon} ^{s+{\rm i}\epsilon} \frac{dz}{z}
\bar{a}^{\rm eff}(-z).
\end{equation}

These equations serve to define the effective coupling
$\bar{a}_s^{\rm eff}(s)$
which parametrizes the $R(s)$ ratio and plays the role of the running
coupling in the timelike region. One usually applies the standard
perturbative approximation for ${a}^{\rm eff}(z)$ to derive the
effective
coupling in the $s$-channel from (\ref{aseff}).
This way leads to the so-called
$\pi^2$-terms which play an important role in the phenomenological
analysis of various processes \cite{pi2}.
However, the perturbative approximation of ${a}^{\rm eff}(z)$
breaks the analytic properties mentioned above. For example, in the
one-loop
approximation the function ${a}^{\rm eff}(z)$ has the form
$1/[\beta_0 \ln(z/\Lambda^2)]$ with a ghost pole at $z=\Lambda^2$,
which contradicts the assumption that ${a}^{\rm eff}(z)$ is an
analytic
function in the cut $z$-plane. A consequence of this problem is
the fact that if ${a}_s^{\rm eff}(s)$, obtained in such
a way, is substituted into (\ref{aeff}), the original one-loop
formula in the spacelike region is not reproduced.

\section{Analytic perturbation theory}

This difficulty can be avoided in the framework of what we call
analytic perturbation theory, in which the running coupling constant
is forced to have the correct analytic
properties.\footnote{The correct analytic properties of the
$D$-function can also be maintained in the framework of the so-called
variational perturbation theory \cite{js} which is based on a new
small
expansion parameter \cite{solov}.}
We define $\bar a^{\rm eff}(q^2)$ in terms of the spectral density
according to (\ref{spectral}), that is, the effective coupling in
the spacelike region is given by
\begin{equation}
\bar a^{\rm eff}(q^2)={1\over\pi}\int_0^\infty{d\sigma\over
\sigma+q^2}\rho(\sigma),
\label{aeffdisp}
\end{equation}
As a result, from (\ref{aseff}) and (\ref{aeffdisp}),
the effective coupling in the timelike
region is given by the following elegant expression:
\begin{equation}
\label{schan}
\bar{a}_s^{\rm eff}(s)\,=\,\frac{1}{\pi }\,
\int_s^\infty\,\frac{d\sigma}{\sigma}\,\,\rho(\sigma)\, .
\end{equation}

It is clear that both coupling constants
$\bar{a}^{\rm eff}(q^2)$ and $\bar{a}_s^{\rm eff}(s)$
have the same universal limit at $q^2=+0$ and $s=+0$ and
a similar tails as $q^2\to \infty$ and $s\to \infty$.
However, in the intermediate region the effect of analytic
continuation
becomes important.  As we will see in perturbation theory, below,
the distinction between the different effective coupling constants
is several percent, which may be important for extracting the QCD
coupling constant from various experimental data.

\section{One-loop results}
Let us consider this problem at the one-loop level.
The perturbative contribution of the leading logarithms to the
effective
coupling in the spacelike region can be written  as follows:
\begin{equation}
\label{aeff2}
\bar{a}^{(1)}(q^2)=a\sum_{n=0}^{\infty}
{\left( -a\beta_0\ln\frac{q^2}{\mu^2}\right)}^n.
\end{equation}
In any finite order this function has the correct analytic
properties.
The ghost pole  appears due to the naive sum of the infinite
geometrical
series in (\ref{aeff2}). However, we should consider the series
in (\ref{aeff2}) as an asymptotic series and try to find its sum
in such a way to maintain the required analytic properties,
taking into account the fact that the sum of an asymptotic series is
not unique. To this end, let us consider the correlation function
 for which, from (\ref{dq2p}) and (\ref{dqeff}),
the contribution of the leading logarithms has the following form:
\begin{equation}
\label{pllog}
\Pi(q^2)\propto -\ln\frac{q^2}{\mu^2}+{\rm const.}
+\frac{d_1}{\beta_0}\sum_{n=0}^{\infty}\frac{1}{n+1}
{\left( -a\beta_0\ln\frac{q^2}{\mu^2}\right)}^{n+1}.
\end{equation}
Carrying out the sum and taking the imaginary part, we immediately
find
from (\ref{rsp})
\begin{equation}
\label{rsarc}
R(s)\propto1+ \frac{r_1}{\pi\beta_0 }{\rm arccos}
\frac{1+a\beta_0 L}{\sqrt{(1+a\beta_0 L)^2+(a\pi\beta_0)^2}},
\end{equation}
where $L=\ln s/\mu^2$. By introducing the QCD parameter
$\Lambda^2=\mu^2\exp(-1/a\beta_0)$, we obtain
for the running coupling constant in the $s$-channel
\begin{equation}
\label{aeffs2}
\bar{a}^{(1)}_s(s)=\frac{1}{\pi\beta_0}\,{\rm arccos}
\frac{\ln (s/\Lambda^2)}{\sqrt{\ln^2( s/\Lambda^2)+\pi^2}}.
\end{equation}
The same result for the running coupling constant in the timelike
region
 can be obtained by substituting the one-loop spectral density
(\ref{specfn})
into (\ref{schan}). Moreover, the substitution of (\ref{aeffs2})
into (\ref{aeff}) reproduces the one-loop analytic running coupling
(\ref{a1}) which parametrizes the $D$-function and has the asymptotic
expansion (\ref{aeff2}). Thus, the summation of the leading
logarithms for the
physical quantity $R(s)$ leads to a $D$-function with
the correct analytic properties.

\begin{figure}
\centerline{
\psfig{figure=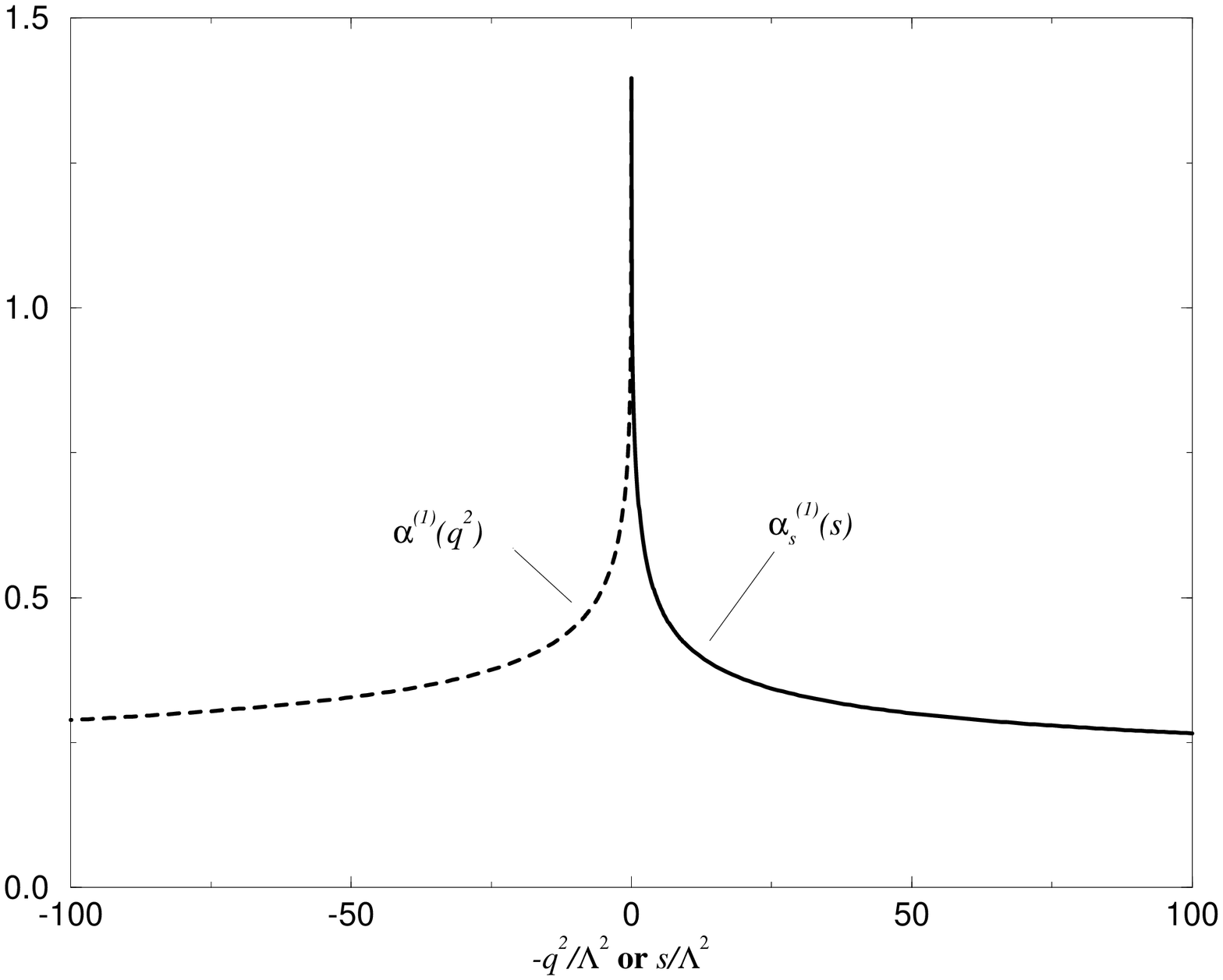,height=5in,width=6.5in,angle=270}}
\caption{Plot of the spacelike and timelike definitions of the
one-loop running coupling
constants, $\bar \alpha^{(1)}(q^2)$ and $\bar \alpha_s^{(1)}(s)$.
The abscissa is
respectively $-q^2/\Lambda^2$ and $s/\Lambda^2$ for the two functions.
Here we have displayed $\alpha=4\pi a$ and have used $N_f=3$ in
$\beta_0$.}
\label{fig1}
\end{figure}

\begin{figure}
\centerline{
\psfig{figure=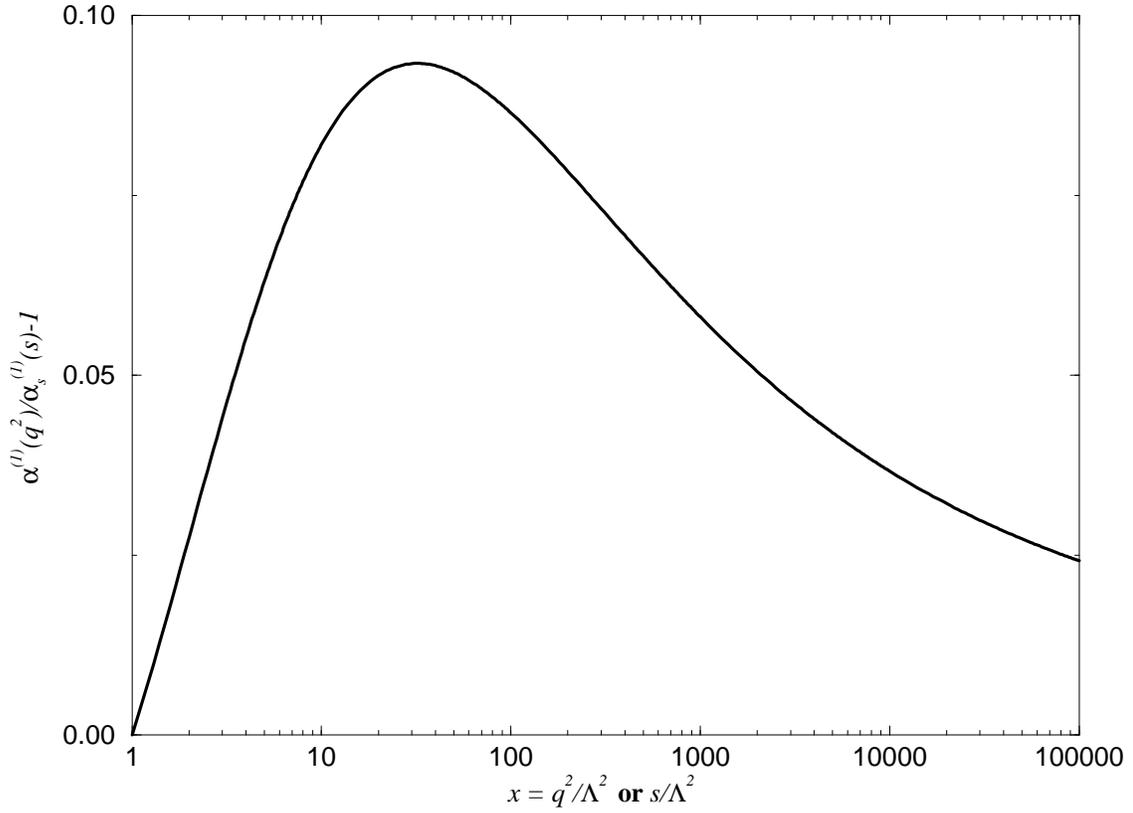,height=5in,width=6.5in,angle=270}}
\caption{Plot of the relative difference of the two coupling
constants
shown in Fig.~\ref{fig1}, $\bar \alpha^{(1)}(x)/\bar
\alpha_s^{(1)}(x)-1$.
Here the argument is $x=q^2/\Lambda^2$ for $\bar \alpha$ and
$x=s/\Lambda^2$
for $\bar \alpha_s$.}
\label{fig2}
\end{figure}

Let us compare the two one-loop couplings $\bar a^{(1)}_s(s)$, given
by
(\ref{aeffs2}), and $\bar  a^{(1)}(q^2)$, given by (\ref{a1}).
As noted above, they have a universal value at 0,
\begin{equation}
\bar a_s^{(1)}(0)=\bar a^{(1)}(0)={1\over\beta_0},
\end{equation} and in fact are exactly the same at $s=\Lambda^2$,
$q^2=\Lambda^2$. Asymptotically, for large spacelike and timelike
momenta, respectively,
\begin{eqnarray}
\bar a^{(1)}(q^2)&\sim&{1\over\beta_0}{1\over\ln q^2/\Lambda^2},\quad
q^2\gg\Lambda^2,\\
\bar a_s^{(1)}(s)&\sim&{1\over\beta_0}{1\over\ln s/\Lambda^2}
\left(1-{\pi^2\over3}{1\over\ln^2 s/\Lambda^2}\right),\quad
s\gg\Lambda^2,
\label{llogs}
\end{eqnarray}
exhibiting the fact that the $t$- and $s$-channel couplings differ in
three-loop
order.  In general, these two couplings, in their respective regimes,
agree numerically quite closely, as shown in Fig.~\ref{fig1}, with
the
relative difference being no more than 9\%, as shown in
Fig.~\ref{fig2}.
Similar features hold in two-loop order, the discrepancy between the
couplings
in the intermediate region dropping to about 5\%.
\section{Schwinger's identification}

More than two decades ago, Schwinger proposed \cite{schwinger}
 that the Gell-Mann--Low function, or the $\beta$-function,
 in QED could be represented by a spectral
function for the photon propagator, which has direct physical
meaning.
The precise connection, of course, depends on the definition of the
running coupling constant \cite{schwinger,milton}.
Remarkably, we find that this idea is realized in our proposal for
the timelike coupling constant in QCD.
 For the $\beta$-function which corresponds to the
coupling defined in the Euclidean region this statement is true
through the
two-loop approximation, but breaks down if one takes into account
three-loop contributions. (The analogous thing happens in QED
for the conventional charge definitions \cite{milton}.)
However, Schwinger's identification is
certainly correct if we construct the $\beta$-function for the
coupling
(\ref{schan})  defined in the timelike region: indeed,
\begin{equation}
\label{betas}
\beta_s=s\frac{d\bar{a}_s^{\rm eff}}{d s}
=-\frac{\rho(s)}{\pi}.
\end{equation}
As noted above,
in perturbation theory, the difference between the couplings in the
spacelike and timelike regions is given by three-loop diagrams
and, therefore,
$\beta=q^2d\bar{a} ^{\rm eff}/dq^2=-\rho(s)/\pi+O(3\mbox{-loop})$.

\section{Conclusions}
We have considered the procedure of constructing the QCD running
coupling constant by using analytic perturbation theory. The
fundamental
quantity here is the spectral density $\rho(\sigma)$, in terms of
which the running coupling in the Euclidean region is expressed
through the spectral representation (\ref{aeffdisp}), while
 the running coupling in the timelike region is expressed by
(\ref{schan}). Both these couplings have the same
universal infrared limit
\begin{equation}
\label{irlimit}
\bar{a}^{\rm eff}(0)=
\bar{a}^{\rm eff}_s(0)=\frac{1}{\pi }\int_0^\infty\frac{d\sigma}
{\sigma}\rho(\sigma)=\frac{1}{\beta_0},
\end{equation}
which turns out to be remarkably stable with respect to higher loop
corrections \cite{ss2}. Further, both  coupling constants have the
same leading
 asymptotic behavior. Thus, in comparison with standard perturbation
 theory, the strong requirement of analyticity modifies the theory
 in the infrared and intermediate domains significantly, which is
 particularly relevant for a physical description of the timelike
 regime. Finally, we have shown that
 Schwinger's proposed connection between the renormalization group
$\beta$-function  and the spectral density is valid for the coupling
defined in the timelike region.

\section*{Acknowledgements}
The authors would like to thank D. V.~Shirkov
for interest in the work and for useful comments.
Partial support of the work by the US National Science Foundation,
grant PHY-9600421, and by the US Department of Energy,
grant DE-FE-02-95ER40923, is gratefully acknowledged.
I. S. also thanks the High Energy Group of the University of
Oklahoma for their warm hospitality.

\bibliographystyle{prsty}

\end{document}